\documentclass[10pt, conference, final]{IEEEtran}
\usepackage{cite}
\usepackage{amsmath,amssymb,amsfonts}
\usepackage{algorithmic}
\usepackage{graphicx}
\usepackage{tabularx}
\usepackage{textcomp}
\usepackage{xcolor}
\usepackage{tikz}
\usepackage{url}

\newcommand{\cmark}[1][]{\tikz[x=1em, y=1em]\fill[#1] (0,.35) -- (.25,0) -- (1,.7) -- (.25,.15) -- cycle;}
\newcommand{\xmark}[1][]{%
\tikz[x=1em, y=1em, line width = .15ex]{
    \draw[line cap=round, #1] (0,0) to[bend left=6] (0.45,0.45);
    \draw[line cap=round, #1] (0,0.45) to[bend right=2] (0.45,0);
    }
}

\pdfoutput=1

\begin{document}
\bstctlcite{IEEEexample:BSTcontrol}

\title{AI-Ready Energy Modelling for Next Generation RAN\\
\thanks{The work of K.~Sthankiya was supported by the UK Engineering and Physical Sciences Research Council (EPSRC) grant {{EP/V519935/1}} for the University of Queen Mary University of London Centre for Doctoral Training in Data-Centric Engineering.}
}

\author{\IEEEauthorblockN{Kishan Sthankiya\IEEEauthorrefmark{1}, Keith Briggs\IEEEauthorrefmark{2}, Mona Jaber\IEEEauthorrefmark{1} and Richard G. Clegg\IEEEauthorrefmark{1}} \IEEEauthorblockA{\IEEEauthorrefmark{1}School of Electronic Engineering and Computer Science\\
Queen Mary University of London, London, UK\\
Email: \{k.sthankiya, m.jaber, r.clegg\}@qmul.ac.uk} \IEEEauthorblockA{\IEEEauthorrefmark{2}Wireless Research, BT Labs, Martlesham Heath, UK\\
Email: keith.briggs@bt.com}}

\maketitle

\begin{abstract}
Recent sustainability drives place energy-consumption metrics in centre-stage for the design of future radio access networks (RAN). At the same time, optimising the trade-off between performance and system energy usage by machine-learning (ML) is an approach that requires large amounts of granular RAN data to train models, and to adapt in near realtime. In this paper, we present extensions to the system-level discrete-event AIMM (AI-enabled Massive MIMO) Simulator, generating realistic figures for throughput and energy efficiency (EE) towards digital twin network modelling. We further investigate the trade-off between maximising either EE or spectrum efficiency (SE). To this end, we have run extensive simulations of a typical macrocell network deployment under various transmit power-reduction scenarios with a range of difference of 43 dBm. Our results demonstrate that the EE and SE objectives often require different power settings in different scenarios. Importantly, low mean user CPU execution times of 2.17 $\pm$ 0.05 seconds (2~s.d.) demonstrate that the AIMM Simulator is a powerful tool for quick prototyping of scalable system models which can interface with ML frameworks, and thus support future research in energy-efficient next generation networks.
\end{abstract}


\section{Introduction}
The significance of mobile telecommunications cannot be overstated, with global subscribers expected to reach 100~billion by 2030~\cite{lopez-perez_survey_2022}. Compared to earlier systems, 5G radio access networks (RAN) have more flexibility to become energy efficient using various techniques~\cite{lopez-perez_survey_2022}. Meanwhile, energy usage remains considerable, with up to 70\% of this concentrated at the base stations (BS)~\cite{piovesan2022machine}. However, exactly how to continuously optimise next generation networks to achieve a trade-off between energy usage and performance is an open question. Artificial Intelligence (AI) and, more specifically, machine learning (ML) techniques are promising for this optimisation, but their training cannot take place on a real network. To enable ML training and testing without the risk of catastrophic network outages, virtual replicas of operational networks mirror events from a live network~\cite{nardini_digital_twin_2022} satisfying the need for vast amounts of real-world data. 


This paper implements and assesses an extension to estimate energy use using a system-level wireless simulator, AIMM (AI-enabled Massive MIMO) Simulator~\cite{AIMM_Simulator}. With our extensions, the AIMM Simulator meets two key needs for using AI to optimise energy usage: rapid scenario analysis for large numbers of training rounds to run and faithful, real-world, network key performance indicator (KPI) tracking; e.g.\ Channel Quality Index (CQI) and Modulation and Coding Scheme (MCS). The simulator is designed to simulate tens of BSs, including multi-tier deployments with tens of User Equipment (UE) devices attached to each. For the scenarios presented in this paper, a single run generally takes less than 3 seconds on consumer hardware. 

Emerging Open RAN~\cite{polese_oran} architecture positions the RAN Intelligence Controller (RIC) as an interface to optimise the network with AI and ML models. By design, the AIMM Simulator offers users an interface to write code in line with the RIC functionality. In other words, integration with the RIC is an abstraction for future development. In this work, we created a function to calculate and monitor the power and energy consumption of base stations in an idealised urban 5G deployment and study scenarios of reduced transmit power. Three-dimensional positions (i.e. $x$, $y$, and $z$ spatial coordinates) of UEs in the simulation area and their CQI and resulting throughput are calculated, allowing a look at realistic estimates for energy efficiency (EE) and spectral efficiency (SE). 

Advancing technology crucially requires dynamic optimisation for resource allocation in rapidly fluctuating network and traffic conditions, where traditional static EE and SE points quickly become obsolete. Our work accentuates this by demonstrating the trade-off between SE and EE and the possibility of major savings by adapting models to the current situation in the network. Minimising the time required to obtain these results and the potential benefits of increasing EE or SE make dynamic optimisation targets crucial for efficient and adaptable network resource utilisation.

The contribution of this paper is presenting the AIMM Simulator and demonstrating its use for measuring energy consumption in different scenarios. The simulator is extremely promising for AI and ML use cases as it can quickly produce a detailed estimate of BS energy consumption in scenarios with tens of BSs and several hundred users. This will enable the deployment of both traditional optimisation techniques and AI training for optimisation techniques that look at, for example, varying power levels and sleep modes. We show that considerable savings can be made in this way, by adjusting only three of the nineteen BS in our scenarios we can increase the network energy efficiency by up to 14.8\%.


\section{Background}
\begin{table*}[!ht]
\renewcommand{\arraystretch}{1.3}
\caption{Comparison of 5G Simulation Tools \label{table:compare_simulators}}
\centering
\begin{tabular}{|r || *{5}{>{\centering\arraybackslash}m{1.3cm}}|}
\hline
\textbf{Feature} & \textbf{AIMM} & \textbf{Py5cheSim} & \textbf{ns3-o-ran-e2} & \textbf{5G-LENA} & \textbf{Vienna}\\
\hline
$\text{3D modelling}$           & \cmark & \xmark & \cmark & \cmark & \cmark    \\
$\text{Free-to-use}$            & \cmark & \cmark & \cmark & \cmark & \xmark    \\
$\text{Open-source}$            & \cmark & \cmark & \cmark & \cmark & \xmark    \\
$\text{Low dependencies}$       & \cmark & \cmark & \xmark & \xmark & \xmark    \\
\hline
\end{tabular}
\end{table*}

An extensive survey~\cite{lopez-perez_survey_2022} for 5G EE, estimates over 50\% of all network energy usage will be the radio access element by 2025. This projection accounts for advanced sleep modes, lean carriers, massive MIMO layer adaption and ML in 5G NR (see~\cite[Section 2]{lopez-perez_survey_2022} for a complete list). At the same time, the report~\cite[Section 5.6]{itu_radiocommunication_bureau_br_future_2022}, envisages AI  as a technology that can anticipate network traffic dynamics and continuously optimise operation. The authors in~\cite{piovesan2022machine} conclude ML techniques may be more promising. However, ML models share issues with all AI models, in that they cannot be trained or tested in operational networks without severe risk of catastrophic outage. Mitigating with virtual representations~\cite{nardini_digital_twin_2022} is a potential solution, but also faces two key challenges when trying to apply ML to optimise energy efficiency of the system. Firstly, they require low latency access to large amounts of reliable performance and energy consumption data, captured with high precision. Secondly, the computational results from the virtual replica must return fast enough to still apply to the physical network. 

For a given network, the energy data captured can be described by an energy consumption model. A common approach is the analytical model in~\cite{holtkamp_parameterized_2013}, estimating that power used by a BS, $P_{\text{BS}} = N_{\text{TRX}} (P_0 + \delta_p P_{\text{out}})$, where $N_{\text{TRX}}$ is the number of transceivers, $P_0$ and $\delta_p$ are power consumption parameters dependent on cell-type and $P_\text{out}$ is the transmit power. Assuming a fixed number of $N_{\text{TRX}}$, we see that power consumption for the whole BS is directly proportional to the transmit power. Later studies model energy use in systems of high and low transmit power nodes across a variety of cloud-RAN (C-RAN) deployment scenarios~\cite{adil_israr_power_2022} and BS sleep modes~\cite{peesapati_analytical_2021}. Each of these works attempts to model energy or power consumption with a specific goal, such as sleep modes~\cite{piovesan2022machine,peesapati_analytical_2021} or network deployment~\cite{adil_israr_power_2022}. However, a lack of realistic parameters is demonstrated by the level of transmit power required to satisfy user demands. This taken from a path loss model following Shannon's theorem~\cite[Eq(2)]{peesapati_analytical_2021} and~\cite[Eq(6)]{adil_israr_power_2022}, which does not account for the required signal strength at the UE to ensure less than 10\% block error rate (BER). Moreover, it is not a focus of these studies to model EE in a network simulator to facilitate fast optimisation in a virtual replica of an operational network.

In isolation, EC alone does not account for how effectively the energy is being used. This is better described by energy efficiency (EE), often expressed as the amount of data transmitted (in bits) per unit of energy consumed (in Joules).
Indeed EE is a complex behaviour that is affected by the network structure and features (see~\cite{lopez-perez_survey_2022}), the stochastic nature of users' spatial and temporal profiles, and environmental conditions. 

In lieu of the first requirement for low latency access to network performance data, described earlier, wireless network simulators provide valuable insights. More specifically, system-level simulators with discrete-event frameworks (see Table~\ref{table:compare_simulators}) are essential for the dynamic modelling of nodes -- reflecting their changes in state and facilitating runtimes which are faster than in real-time. However, there is diversity in the tools available to model next generation systems in the literature. For example, Py5cheSim~\cite{pereyra_py5chesim_2021} falls short in capturing features such as the simulation nodes having $x$, $y$, and $z$ spatial coordinates (\emph{3D~modelling}). Use of these co-ordinates are crucial for predicting service interruptions and user experience degradation for key enabling technologies in 5G, such as massive MIMO and beamforming operations.
Meanwhile, efforts to support open innovation can be fostered through transparent practices, such as \emph{open-source} and \emph{free-to-use} software. Most simulators share this trait, yet the Vienna~\cite{muller_vienna_2018} project runs in MATLAB which is proprietary and incurs licensing fees. 
Tools with many dependencies necessitate a larger codebase, increasing complexity and crucially introducing longer runtimes for a virtual replica. Amongst the ns-3 based, 5G-LENA~\cite{mezzavilla_5G_LENA_2018} and ns3-o-ran-e2~\cite{lacava_ns3_oran_e2_2023}, the former requires four supplementary libraries and the latter requires two. Given that virtual replicas require the ability to evaluate many scenarios with incredible speed, applying an upper limit of three external libraries, the AIMM Simulator~\cite{AIMM_Simulator} becomes a clear contender, only requiring Python 3.8+ and two additional libraries with the entire codebase standing at $\leqslant$ 10MB.

It is evident that there is a lack of lightweight network simulation tools that dynamically model energy efficiency towards facilitating fast optimisation in data-driven virtual replicas. To address the challenges of providing accurate energy data that mirror real-world conditions at low latencies and support open research efforts, the AIMM Simulator is positioned as prominent solution. These findings indicate a timely opportunity to expand the capabilities of the AIMM Simulator, by building an extension for energy modelling. This extension bridges the gaps in the leading 5G network simulation and energy modelling research to enable the potential of ML techniques to optimise energy efficiency in operational networks. It is described in section~\ref{sec:power}.

\section{Methods}
\subsection{AIMM Simulator} 
\begin{figure}[!ht]
\centering
\includegraphics[width=3.0in]{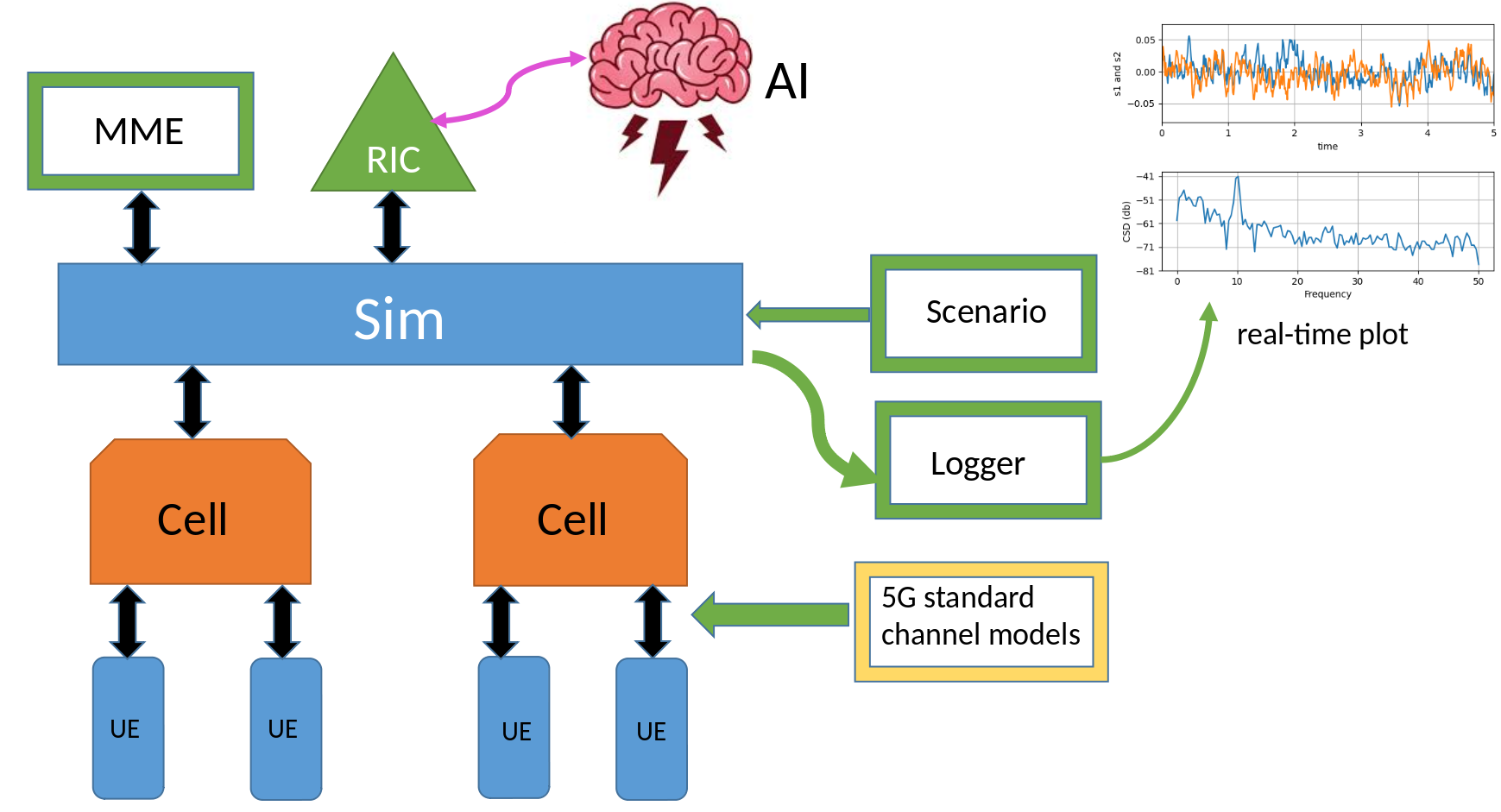}
\caption{AIMM Simulator Block Diagram~\cite{AIMM_Simulator}}
\label{fig:aimm}
\end{figure}
The AIMM Simulator is a fast system-level simulator developed to recreate 5G New Radio (NR) elements as seen in \figurename~\ref{fig:aimm}. Written in Python, the core simulator (Sim) employs a discrete-event library for fine control over component interaction. Physical and network layer concepts (e.g.\ resource elements, IP packets) are abstracted to reduce runtime. Finally, numerical Python enhances operation speed comparable to compiled C code. 

The flexibility of simulations is first demonstrated within a Sim instance, where a central loop interval orchestrates downstream events for attached components. BSs (Cell) and user equipment (UE) are created in a 3D space with the option to override default parameters (see Table~\ref{table:params}). 
A Mobility Management Entity (MME) maintains UE attachment and handover, where strategies and link-level threshold logic can be defined. More sophisticated control is reserved for the RIC, which provides a `clean' interface to allow integration with ML libraries (e.g.\ TensorFlow, PyTorch) and development of rApps and xApps. Dynamic features (e.g.\ cell transmit power) may be customised specifically to an experiment via the Scenario class, and polling the state of the system is handled by the Logger. These features highlight the tool's versatility and foster a firm grounding for evaluating EE and SE in next generation networks.

The AIMM Simulator features standardised pathloss models~\cite{3rd_generation_partnership_project_study_2017}, namely: Urban Macro (3D-UMa), Urban Micro (3D-UMi), and  Indoor Hotspot (3D-InH) --- including both line-of-sight (LoS) and non-line-of-sight (NLoS) variants. 
These models are used to calculate Reference~Signal~Received~Power (RSRP) and Signal-to-Interference-plus-Noise~Ratio (SINR):
\begin{IEEEeqnarray}{rCl}
    \text{RSRP}_{i,j} & = & G^{\text{MIMO}}_{j} + G^{\text{ant}}_{j} + P^{\text{Tx}}_{j} - \text{PL}_{i,j}
    \label{eq:rsrp} \\
    \text{SINR}_{i,j} & = & \frac{{P^{\text{Rx}}_{i,j}}}{{P^{\text{inter}}_{i} + P^{\text{noise}}_{i}}}
    \label{eq:sinr}
\end{IEEEeqnarray}
where indices $i$ and $j$ refer to the UE and the BS, respectively. The following parameters are used:
\begin{table}[htbp]
\centering
\begin{tabular}{lp{6cm}}
    $G^{\text{MIMO}}$ & MIMO (Multiple-Input Multiple-Output) gain, \\
    $G^{\text{ant}}$ & Antenna gain, \\
    ${P}^{\text{Tx}}$ & Transmit power, \\
    $\text{PL}$ & Pathloss between BS and UE, \\
    $P^{\text{Rx}}$ & Received power of the signal from serving BS, \\
    $P^{\text{inter}}$ & Interference from other BS experienced by UE, \\
    $P^{\text{noise}}$ & Noise power at the UE. \\
\end{tabular}\label{table:rsrp-sinr}
\end{table}


Unlike traditional approaches that rely on Shannon's theory to estimate the user throughput, AIMM uses pre-computed tables to look up CQI values based on SINR values at 10\% Block Error-Rate thresholds. 
Linear scaling of CQI to MCS index allows lookup of SE from~\cite[Table 5.1.3.1-1]{3GPP_ts38214_rel17}.
It follows that the throughput of the $i$-th UE attached to the $j$-th BS is:
\begin{IEEEeqnarray}{rCl}
    {T}_{i,j} & = & \text{SE}_{i,j} \times B \text{,}
    \label{eq:tp_ue_cell_subband}
\end{IEEEeqnarray}
where $\text{SE}_{i,j}$ represents the SE, determined by the MCS index (obtained from the SINR in (\ref{eq:sinr})) and $B$ is the bandwidth.

\subsection{Extension Implementing Power Consumption Model for Base Stations}
\label{sec:power}
The power consumption of BSs is estimated based on a model inspired by~\cite{holtkamp_parameterized_2013}. The total power consumption of the $j$-th BS is denoted as $P^{\text{BS}}_{j}$,
\begin{IEEEeqnarray}{rCl}
    P^{\text{BS}}_{j} & = & {N}_{\text{TRX}} \times {N}_\text{ant} \times ({P}_{\text{0}} + f(P^{\text{Tx}}_j))\text{,}
    \label{eq:overall_BS_power}
\end{IEEEeqnarray}
and is the product of the number of transceiver chains (${N}_{\text{TRX}}$), number of antennas (${N}_\text{ant}$) and the static power consumption that is not related to the actual load of the BS ($P_{\text{0}}$) plus a function of the radiated transmit power ($P^{\text{Tx}}_j$) of the $j$-th BS, where
\begin{IEEEeqnarray}{rCl}
    f(P^{\text{Tx}}_j) & = & \frac{\frac{P^{\text{Tx}}_j}{\eta_{\text{PA}} \cdot \left(1-\sigma_{\text{feed}}\right)} + P_{\text{RF}} + P_{\text{BB}}}
    {\left(1-\sigma_{\text{DC}}\right) \left(1-\sigma_{\text{MS}}\right) \left(1-\sigma_{\text{cool}}\right)}\text{.}
    \label{eq:BScomponents}
\end{IEEEeqnarray}
Refer to Table~\ref{table:params} for the definitions and values of the variables in this model.

To add energy modelling capabilities, an extension module was created for the AIMM Simulator which can be found on GitHub~\cite{AIMM_EnergyModels}. This module includes a system power consumption model and parameters for different types of BS~\cite{holtkamp_parameterized_2013}, 5G CQI reference tables~\cite{3GPP_ts38214_rel17}, cell sleep mode, per-user SINR based handover and logging for later analysis. Furthermore, we define experimental scenarios to allow tracking of UE movements and changes to transmit power of BSs. 

\subsection{System Model}
\begin{figure}[!ht]
\centering
\includegraphics[width=3.5in]{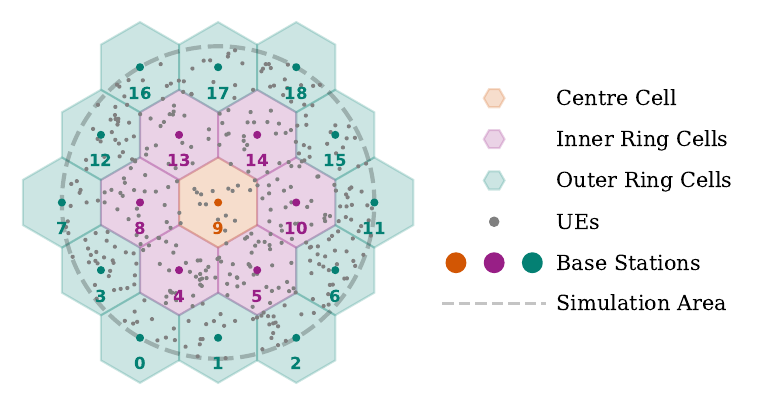}
\caption{Topology of our system with nineteen BSs in a regular grid and UEs deployed over this grid.}\label{fig:sys}
\end{figure}

We consider the downlink of a homogeneous macro cell RAN to explore the energy efficiency. As seen in \figurename~\ref{fig:sys}, we deploy macro BSs, in a regular hexagonal grid, with centre frequency $\nu$ and bandwidth $B$. BSs are evenly spaced with an inter-site distance $D_\text{isd}$ and height ($H_\text{BS}$). Each BS serves the geographical area of a BS numbered in \figurename~\ref{fig:sys}, composed of transceivers ($N_\text{TRX}$) each served by a number of antennas ($N_\text{ant}$). 
At each antenna, we assume an omnidirectional radiation pattern and a transmit power in Watts ($P_j$), where $P_j \leqslant P_\text{max}$, the maximum transmit power. We group BSs by position, described in Table~\ref{table:cell_scenarios}.%

UEs ($i= 1,2, \ldots, {N}_\text{UE}$) are distributed with a homogeneous Poisson point process with density $\lambda_\text{UE}$ UEs per $\text{km}^2$. UE height is defined as ${H}_\text{UE}$ and is the same for all. UEs are assumed to be outdoor and stationary, with a 3D-UMa NLoS pathloss model with serving and neighbouring BSs. The operating mode is frequency-division duplexing (FDD), and UEs attach each to a maximum of one BS and select the BS with the highest SINR based on (\ref{eq:sinr}). The user-to-BS connectivity is represented by defining in $X_{i,j}$ for each $\text{UE}_{i}$ and each $\text{BS}_{j}$:
\begin{IEEEeqnarray}{rCl}
    X_{i,j} & = & \begin{cases}
      1 & \text{if $\text{UE}_{i}$ is connected to $\text{BS}_{j}$}\\
      0 & \text{otherwise}
    \end{cases}
    \label{eq:ser}
\end{IEEEeqnarray}
This power consumption model provides an estimation of the energy consumed by the BSs, taking into account various components and efficiency factors.

\begin{table}[htbp]
\caption{Definition of variables in the power consumption model}
\centering
\begin{tabular}{llr}
\hline
\textbf{Variable} & \textbf{Description} & \textbf{Value} \\
\hline
$\nu$&Carrier frequency  & 3.5 GHz \\
${B}$&BS bandwidth  & 10 MHz \\
${N}_\text{BS}$&No. of BS  & 19 \\
${H}_\text{BS}$&BS height & 25 m \\
${{P}_\text{max}}$&Max transmit power & 20 W \\
${D}_\text{isd}$&BS Inter-site distance & 500 m \\
$\lambda_\text{UE}$ & UE density & 1256 per km\textsuperscript{2}\\
${H}_\text{UE}$& UE height & 1.5 m \\
${N}_{\text{TRX}}$ & No. of transceivers in the base station & 6\\
${N}_{\text{ant}}$ & No. antennas & 1\\
${P}^{\text{Tx}}$ & Base station radiated transmit power & 0--20 W\\
${P}_{\text{0}}$& Static power & 130 W\\
$\eta_{\text{PA}}$ & Power amplifier efficiency & 0.311\\ 
${P}_{\text{RF}}$ & RF processing power & 12.9~W\\
${P}_{\text{BB}}$ & Baseband processing power & 29.6~W\\
$\sigma_{\text{feed}}$ & Feeder loss & 0.5 \\
$\sigma_{\text{DC}}$ & DC-to-AC losses & 0.075 \\
$\sigma_{\text{MS}}$ & Mains supply losses & 0.09\\
$\sigma_{\text{cool}}$ & Cooling losses & 0.10\\ [1ex]
\hline
\end{tabular}\label{table:params}
\end{table}
\begin{table}[htbp]
\caption{BS Scenario Definitions (refer to \figurename~\ref{fig:sys})}
\centering
\begin{tabular}{l l r}
\hline
\textbf{Scenario} & \textbf{Description} &\textbf{Variable Power BSs} (K\textsuperscript{v}) \\
\hline
$\text{Scenario}~\text{1}$ & Centre & 9  \\
$\text{Scenario}~\text{2}$ & Inner ring, antipodal & 8, 10\\
$\text{Scenario}~\text{3}$ & Inner ring, alternate & 4, 10, 13 \\
$\text{Scenario}~\text{4}$ & Central triad & 4, 8, 9 \\
\hline
\label{table:cell_scenarios}
\end{tabular}
\end{table}
\subsection{Experiment Setup}\label{sec:Exp}
We begin by varying transmit power (${P}^{\text{Tx}}_j$) of different BSs within the network in \figurename~\ref{fig:sys}, implementing scenarios as defined in Table~\ref{table:cell_scenarios}. In each scenario, the BSs in~\figurename~\ref{fig:sys} are separated into two sets: 
$K^\text{v}$ the variable power BSs and $K^\text{s}$ fixed power BSs, such that $\mid K^{\text{v}} \cup K^\text{s} \mid=J$.

In each of the four scenarios in Table~\ref{table:cell_scenarios}, we study the effects of reducing the transmit power ($P^\text{Tx}_j$), from ${P}_\text{max}$ to 0 W (${P}_\text{sleep}$) in steps of 3 dBm on multiple metrics. In total, 1600 simulation runs are carried out per scenario (i.e. 100 seed values per power level). We first capture the throughput as defined in (\ref{eq:tp_ue_cell_subband}) and power consumption as defined in (\ref{eq:overall_BS_power}) for each of the ${J}$ BSs. The mean throughput of $\text{BS}_{j}$ is:
\begin{IEEEeqnarray}{rCl}
    \bar{T}_j & = & \frac{\sum_{i=1}^{{N}_{\text{UE}}}\ {T}_{i,j}\times X_{i,j}}{\lvert \sum_{i=1}^{{N}_{\text{UE}}}\ X_{i,j} \rvert}\text{,}
    \label{eq:Tbar}
\end{IEEEeqnarray}
where $X_{i,j}$ indicates if the $i$-th UE is connected to the \mbox{$j$-th} BS and $T_{i,j}$ is defined in (\ref{eq:tp_ue_cell_subband}). Based on the \textit{BS-based} metrics defined in (\ref{eq:tp_ue_cell_subband}), (\ref{eq:overall_BS_power}), and (\ref{eq:Tbar}), we formulate four \textit{set-based} metrics, where a set $S$ could be $K^\text{v}$, $K^\text{s}$, or $K^\text{v} \cup K^\text{s}$.
\begin{itemize}
    \item Mean throughput (Mb/s) per set ($T_{S}$) as in (\ref{eq:Tk}),
    \item Mean power consumption (kW) per set ($\text{PC}_{S}$) as in (\ref{eq:PCk}),
    \item Mean spectrum efficiency (Mb/s/Hz): $\text{SE}_{S}=T_{S}/B$ where $B$ is the bandwidth,
    \item Mean energy efficiency (Mb/J): ${T_{S}\times \tau}/{\text{PC}_{S}}$, where $\tau$ is the duration of one run in seconds where,
\end{itemize}
\begin{IEEEeqnarray}{rCcR}
    T_{S} & = & {\frac{\sum \bar{T}_S}{\mid S \mid}} \quad &\forall  \quad  \text{BS}_{j} \in  S \text{,}
    \label{eq:Tk} \\
    \text{PC}_{S} & = & {\frac{\sum P_{\text{BS}_{j}}}{\mid S\mid}} \quad &\forall \quad \text{BS}_{j} \in  S \text{.}
    \label{eq:PCk}
\end{IEEEeqnarray}

Each of the four scenarios in Table~\ref{table:cell_scenarios} is repeated for 100 different seed values and the metrics are averaged over all runs per scenario. In each scenario, the performance metrics for three sets are calculated $K^\text{v}$, $K^\text{s}$, and $K^\text{v} \cup K\textsuperscript{s}$. 

\begin{figure*}
  \centering
  \includegraphics[keepaspectratio=True, width=\textwidth]{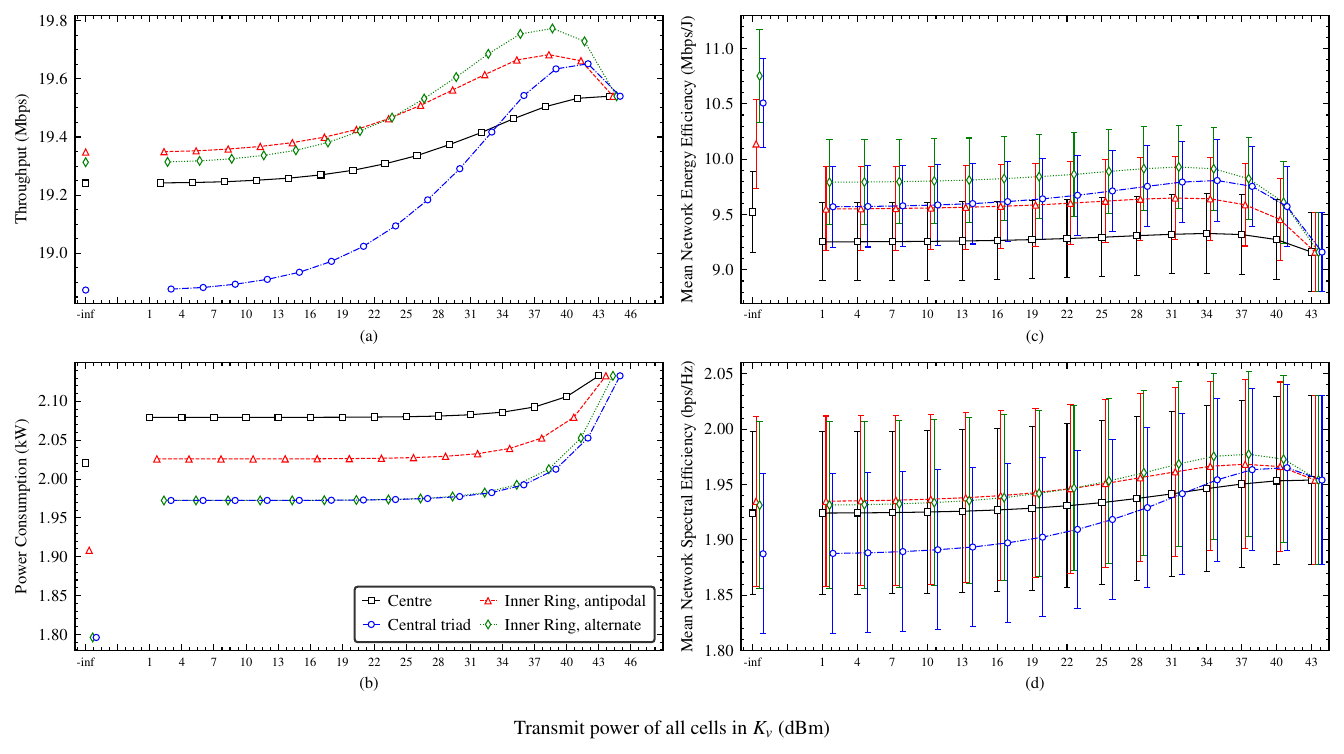}
  \caption{Comparison of Mean Network Throughput (a), Mean Network Power Consumption (b), Mean Network Energy Efficiency (c) and Mean Network Spectral Efficiency (d). Each subplot $x$-axis represents the $P^\text{Tx}_j \  \forall$ K\textsuperscript{v} , ranging from sleep mode ($-\text{inf}$~dBm) to 43~dBm. All plots illustrate the effect of reducing transmit power of BSs in ~(K\textsuperscript{v}), and the impact that has on the network mean (i.e., across all cells).}
  \label{fig:tp_ec_ee_se}
\end{figure*}

\section{Results and Discussion}
In this section, we present the results of implementing the scenarios in Table~\ref{table:cell_scenarios}, as shown in \figurename~\ref{fig:tp_ec_ee_se}. For each scenario, we show the mean network throughput (as in (\ref{eq:Tk}), plot (a) and mean network power consumption (as in (\ref{eq:PCk}), plot (b). Next, the mean values of EE and SE in the whole network for each scenario are presented in (c) and (d). In the following sections, we discuss the impact of power setting in each scenario based on these four metrics.

\subsection{Impact on throughput and power consumption}
The network throughput in \figurename~\ref{fig:tp_ec_ee_se}(a) is affected by power $P^\text{Tx}_j$, with lower output power resulting in lower throughput. However, the peak throughput for each scenario occurs at different power settings. For example, reducing the transmit power of the \emph{Inner Ring, alternate} has a greater impact on throughput than \emph{Central Triad}, despite similar power consumption profiles and the same number of cells. It follows that optimising the choice of cells to reduce power and the setting of $P^\text{Tx}_j$ is crucial for achieving the highest network throughput with the lowest energy consumption.

In plot (a), it is evident that reducing $P^\text{Tx}_j$ by 6~dBm leads to an increase in throughput, except in the \emph{Centre} scenario. This increase in throughput is explained by (\ref{eq:sinr}), where a reduction in unwanted signals from nearby BSs leads to an increase in both SINR and CQI. Therefore, this change in output power proves to be an effective solution for mitigating interference and improving the system's overall performance in most scenarios.

\subsection{Energy and Spectrum Efficiency}
This section evaluates the network's EE across different scenarios, as shown in \figurename~\ref{fig:tp_ec_ee_se}(c). The EE enhancements range from 34-43 dBm, with the most significant gains observed in the \emph{Inner Ring, alternate} setup, resulting in a 0.768 Mbps/J boost. However, the EE decreases slightly when the base station (BS) power drops below these levels in all scenarios. Notably, the EE significantly improves when the BSs in K\textsubscript{v} are in sleep mode (-inf). The \emph{Inner Ring, alternate} setting yields the most substantial EE improvements of 14.8\%, while the \emph{Centre} scenario shows the lowest improvement at 3.79\%. These highlight the importance of optimal strategy selection for reducing network power and its implementation timing.

Reducing the transmit power of a BS ($P^\text{Tx}_j$) can decrease the network's overall SE. In \figurename\ref{fig:tp_ec_ee_se}(d), we observed a slight increase in SE when $P^\text{Tx}_j$ was reduced from $P_\text{max}$ to 37 dBm. However, when all active BSs operate at maximum capacity, less power is available for data transmission per unit of time, resulting in decreased SE. For instance, the \emph{Central Triad} scenario in our analysis experienced a 3.9\% reduction in SE performance. On the other hand, the \emph{Centre} scenario had only a minimal impact on SE, causing a 1.5\% reduction. Moreover, the \emph{Inner Ring, alternate} scenario had better SE performance in the range of $P_\text{max}$ to 22 dBm. In contrast, the \emph{Inner Ring, antipodal} performed better at lower $P^\text{Tx}_j$ levels. Therefore, reducing BS $P^\text{Tx}_j$ can have varying impacts on SE, depending on the specific scenario. The error bars in \figurename \ref{fig:tp_ec_ee_se}(c) and (d) represent one standard deviation from the mean, primarily due to the static placement algorithm and the influence of UE placement on both EE and SE.

\subsection{Lessons learnt for future network design}
The findings of this study underscore the intricate nature of jointly optimising energy and spectrum efficiency, necessitating meticulous selection of BSs and precise configuration of power levels. For instance, in the context of pursuing pure EE objectives, the \emph{Inner Ring, alternate} scenario emerges as the optimal choice when three BSs in K\textsubscript{v} are in sleep mode. Conversely, when the objective shifts to pure SE, the same scenario achieves peak performance with an output power of $P^\text{Tx}_j=37~\text{dBm}$. These instances exemplify the nuanced trade-offs between EE and SE goals, necessitating a judicious balance in optimisation based on network metrics and user requirements.

The research further highlights the AIMM Simulator's potential to model interactions between EE and SE and its capacity for realistic throughput, in contrast to Shannon-based estimations. The operational mode of FDD supports energy-efficient carrier shutdown methods, with strategies such as prolonged periods of deep sleep lasting minutes or hours. Strategies like cell zooming enable a macro cell's coverage expansion while allowing others to remain in a deeper sleep for extended periods~\cite{lopez-perez_survey_2022}, ensuring efficient energy usage without compromising network performance.

Additionally, the study underscores the practical applicability and potential advantages of integrating the AIMM Simulator into AI training and real-world deployments. Analysis of 100 user CPU execution times reveals a mean of 2.17 $\pm$ 0.05 seconds (2~s.d.) per simulation run, achieved without specialised hardware\footnote{Simulations were run on a M1 Macbook Air (Model:Z1250001QB/A).}. This outstanding speed positions the AIMM Simulator as a promising tool for evaluating AI applications in the context of a digital twin and within operational live networks.

\section{Conclusions}
In this work, we present a study on RAN efficiency under different scenarios of BS power settings. To this end, we first describe the adapted energy consumption model and its use in the AIMM Simulator. We define a system of 19 BSs and formulate four scenarios for manipulating the transmit power levels. Our results shed light on the interdependence between energy and spectral efficiency in a realistic representation of throughput estimation. Our findings emphasise the practical application and benefits of utilising the AIMM Simulator for fast and lightweight simulations. Furthermore, given the performance, we believe the AIMM Simulator is an effective AI-ready tool for yielding fast results as a virtual replica for next generation mobile networks.

\section*{Acknowledgment}

 Keith Briggs developed the original AIMM Simulator code as part of the AIMM project\footnote{\url{https://www.celticnext.eu/project-aimm/}}. The full source code is available at \url{https://github.com/keithbriggs/AIMM-simulator}. \mbox{Kishan}~Sthankiya developed extensions for energy modelling.

\enlargethispage{-0.94in}
\bibliographystyle{IEEEtran}
\bibliography{IEEEabrv,mybibfile}

\end{document}